\documentclass[a4paper, pra, twocolumn,superscriptaddress,showpacs,english]{revtex4}
\usepackage[T1]{fontenc}
\usepackage[latin9]{inputenc}
\usepackage{float}
\usepackage{amssymb}
\usepackage{amsmath}
\usepackage{graphicx}
\usepackage{babel}
\usepackage{color}
\usepackage[colorlinks,citecolor=blue]{hyperref}

\begin{document}

\title{Bose-Einstein condensates in an eightfold symmetric optical lattice}

\author{Zhen-Xia Niu}
\affiliation{Department of Physics, Renmin University of China,
Beijing 100872, China}
\author{Yonghang Tai}
\affiliation{Yunnan Key Laboratory of Optoelectronic Information Technology, Yunnan Normal University, Kunming 650500, Yunnan, China}
\author{Junsheng Shi}
\affiliation{Yunnan Key Laboratory of Optoelectronic Information Technology, Yunnan Normal University, Kunming 650500, Yunnan, China}
\author{Wei Zhang}
\email{wzhangl@ruc.edu.cn}
\affiliation{Department of Physics, Renmin University of China,
Beijing 100872, China}
\affiliation{Beijing Key Laboratory of
Opto-electronic Functional Materials and Micro-nano Devices,
Renmin University of China, Beijing 100872, China}

\begin{abstract}
We investigate the properties of Bose-Einstein condensates (BECs) in a two-dimensional
quasi-periodic optical lattice (OL) with eightfold rotational symmetry by numerically solving the Gross-Pitaevskii equation. 
In a stationary external harmonic trapping potential, we first analyze the evolution of matter-wave interference pattern 
from periodic to quasi-periodic as the OL is changed continuously from four-fold periodic and eight-fold quasi-periodic.
We also investigate the transport properties during this evolution for different interatomic interaction and lattice depth, 
and find that the BEC crosses over from ballistic diffusion to localization. Finally, we focus on the case of eightfold 
symmetric lattice and consider a global rotation imposed by the external trapping potential. 
The BEC shows vortex pattern with eightfold symmetry for slow rotation, becomes unstable for intermediate rotation, 
and exhibits annular solitons with approximate axial symmetry for fast rotation. These results can be readily demonstrated
in experiments using the same configuration as in Phys. Rev. Lett. 122, 110404 (2019).

\end{abstract}

\maketitle

%%%%%%%%%%%%%%
%%%%%%%%%%%%%%
\section{Introduction}

Quasicrystals, which can exhibit long-range order without
translational symmetry, have been used to study quantum states
between the limits of periodic order and disorder.~\cite{copyQC1,copyQC2,copyQC3}
According to the theorems of crystallography, the rotational symmetries of periodic lattices
are highly restricted to a few possibilities, namely two-, three-, four-, and
six-fold symmetries. However, quasicrystals can show all the
rotational symmetries forbidden to crystals, including five-,
seven-, eight-, and higher-fold symmetries,~\cite{copyQC,copyEX} 
as can be retrieved from experimental data of electron microscopy, 
spectroscopy, surface imaging methods, and diffraction pattern.~\cite{copyEX}

The remarkable tunability and dynamical control of optical lattices
(OLs) offer an opportunity to investigate quantum many-body states
in periodic systems.~\cite{copyOL1,copyOL2} Ultracold atomic gases
trapped in OLs can work as promising candidates to simulate versatile 
quantum phenomena in various fields of physics, such as condensed matter
physics, high energy physics, and astrophysics. For example, the novel 
quantum transition from extended to localized states has been 
investigated in numerical simulation and observed in experimental
exploration in the presence of OLs.~\cite{copySFL1,copySFL2,copySFL3,copySFL4}
In addition, a carefully arranged configuration of laser beams can 
generate a quasi-periodic optical lattice with fascinating spatial
patterns which are neither periodic as crystals (i.e. lack of
translational symmetry) nor totally disordered (i.e. possession of
long-range order).~\cite{copyQOL1,copyQOL2,copyQOL3,copyQOL4}
Recently, a Bose-Einstein condensate (BEC) has been realized experimentally 
in a two-dimensional (2D) quasi-periodic OL with eightfold rotational symmetry.~\cite{copyEight} 
This achievement paves the route to study quasi-periodic systems which 
are expected to present hybrid features of crystals and amorphous matters.

Stimulated by the experimental achievement of BECs in eightfold symmetric 
optical lattices,~\cite{copyEight} we study in this work the effects of quasi-periodicity on various 
static and dynamic properties of BECs. By numerically solving the Gross-Pitaevskii (GP)
equation, we first discuss the matter-wave interference pattern of BECs 
obtained from time-of-flight images, which reveals the eightfold symmetry 
and the self-similarity of the quasicrystal lattice. 
Next, we investigate the diffusion of particles upon released from the 
harmonic trap while the background lattice potential is present. A crossover 
from the ballistic diffusion to spatial localization is observed by continuously 
tuning the lattice configuration from a periodic square lattice to an eightfold 
symmetric quasi-periodic lattice. The effects of interaction and lattice depth 
on particle diffusion are also discussed. In addition, by applying a 
tilt of the lattice potential, we study the quasi-periodic Bloch oscillation
and compare to the conventional Bloch oscillation induced by
static force in a regular OL. Finally, we impose an external rotation on
the system and numerically solve for the ground state of BECs in the 
combined potential of OL and harmonic trap. When the rotation is slow in comparison
to the harmonic trapping frequency, vortices can be generated to form a 
lattice structure with eightfold symmetry. When the rotating frequency is much 
greater than the trapping frequency, annular solitons can be formed along 
the radial direction with an approximate axial symmetry. For intermediate 
rotating speed, the system is dynamically unstable with no steady solution.
These results extend our understanding on BECs through the crossover from 
ordered to disordered lattices, and can be readily implemented in experiments.

The remainder of this manuscript is organized as follows. In Sec.~\ref{sec:Model} we 
introduce the configuration of the system under consideration and present the Gross-Pitaevskii equation.
We then solve the GP equation with no external rotation and discuss in Sec.~\ref{sec:transport} 
the matter-wave interference pattern, particle diffusion, and Bloch oscillation
in the presence of quasi-periodic lattices. By imposing an overall rotation with different
frequencies, we study the emergence and structure of vortices and solitons in Sec.~\ref{sec:vortices}. 
Finally, we summarize the main results in Sec.~\ref{sec:con}.

%%%%%%%%%%
\section{Formalism}
\label{sec:Model}

%%%%%%%%%%%%%%%%%%%%%%%%%%%%%%%%%%%%%%%%%%%%%%%%%%%%%%%%%%%%%%%%%%%
\begin{figure}[tbp]
\begin{center}
\rotatebox{0}{\resizebox *{5.5cm}{9.0cm} {\includegraphics
{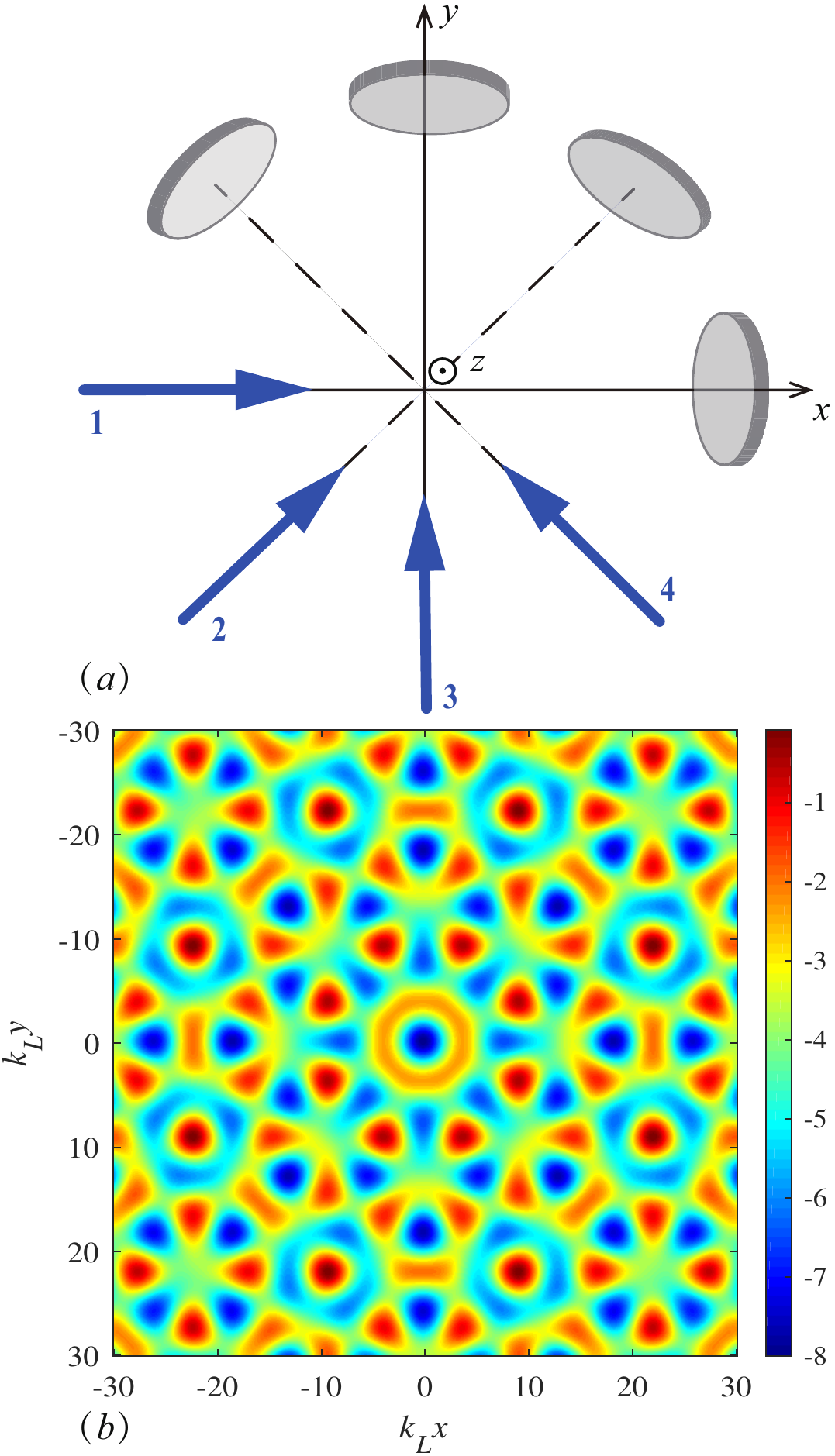}}}
\end{center}
\caption{(Color online) (a) Schematic illustration of laser arrangement to
create an eightfold symmetric optical lattice. (b) A false-color plot of the 
quasi-periodic lattice potential with $V_{0}=-2$.} \label{fig:schematic}
\end{figure}
%%%%%%%%%%%%%%%%%%%%%%%%%%%%%%%%%%%%%%%%%%%%%%%%%%%%%%%%%%%%%%%%%%%

The structure of a two-dimensional eightfold rotationally symmetric lattice
can be created by four mutually incoherent one-dimensional (1D) optical lattices, 
which are yielded by retro-reflection of four single frequency laser beams,
lying in the $x$-$y$ plane and separated by an angle of $\pi/2$. 
The four lasers are all linearly polarized to be perpendicular to the plane
as illustrated in Fig.~\ref{fig:schematic}(a).~\cite{copyEight}
In general, the resulting potential of such a laser configuration can be 
expressed as
\begin{equation}
\label{eqn:1}
%\begin{split}
V_{\rm latt}(\textbf{r}) = V_{0} \sum_{i=1}^{4} \varepsilon_{i} 
\cos^{2}\left(\frac{\textbf{\^{G}}_{i}}{2}\cdot\textbf{r}\right),
%\end{split}
\end{equation}
where $\textbf{r}=(x,y)$ labels the 2D spatial coordinate in the lattice plane, 
$0\leq\varepsilon_{i}\leq1$ stands for the relative dimensionless
intensity of the $i^{\rm th}$ laser beam, and $V_0<0$ is the overall lattice depth. 
The reciprocal lattice vectors $\textbf{\^{G}}_{i}$ of the
four 1D lattices are arranged as $\textbf{\^{G}}_{1} \propto (1,0)$,
$\textbf{\^{G}}_{2} \propto (1,1)/\sqrt{2}$, $\textbf{\^{G}}_{3} \propto (0,1)$
and $\textbf{\^{G}}_{4} \propto (-1,1)/\sqrt{2}$, respectively.
The magnitude of the lattice vectors are normalized as $|\textbf{\^{G}}_{i}|=2 \pi k_{L}$, 
where $k_{L}$ is the lattice wave vector. In the following discussion, we set the 
lattice wavelength $\lambda=2\pi/k_{L}$ as the length unit, the recoil energy 
$E_{R}=\hbar^{2}k_{L}^{2}/2m$ with $m$ the atomic mass as the energy unit,
and the recoil frequency $\omega_{R}=E_{R}/\hbar$ as the unit of frequency. 
With this configuration, one can tune the lattice potential continuously from a 2D cubic lattice with 
$\varepsilon_1 = \varepsilon_3 = 1$ and $\varepsilon_2 = \varepsilon_4 = 0$ 
to an eightfold symmetric quasi-periodic lattice with all $\varepsilon_{i} = 1$. 
In Fig.~\ref{fig:schematic}(b), we show a false color plot of the lattice potential for 
the latter scenario with eightfold symmetry for $V_0 = - 2 E_R$. The quasi-periodic lattice 
shows no translational invariance but some apparent long-range order. 

We consider interacting BECs trapped in a combined potential
of the lattice potential Eq.~(\ref{eqn:1}) and a cylindrically symmetric 
external harmonic trap $V_{\rm trap} = m \omega^2 r^2/2 + m \omega_z^2 z^2/2$. 
In the quasi-2D regime where the trapping frequency along the $z$-axis $\omega_z$ 
is much greater than that in the radial plane $\omega$, the atoms are tightly confined 
along $z$-direction such that only the ground harmonic oscillator is occupied. 
Thus, one can integrate out the degrees of freedom along the $z$-axis,
and consider a 2D system trapped in a combined potential 
\begin{equation}
\label{eqn:2}
\begin{split}
V(\textbf{r})=V_{\rm latt}(\textbf{r})+\frac{1}{2} \omega^{2}r^{2}.
\end{split}
\end{equation}
Here, we have employed the proper units to obtain a dimensionless form.

The dynamical evolution of the BEC wave function $\psi(\textbf{r},t)$ in such a potential can be described 
by the time-dependent Gross-Pitaevskii equation, which takes the following dimensionless form
\begin{equation}\label{eqn:3}
\begin{split}
i\frac{\partial\psi(\textbf{r},t)}{\partial t}
=\left[-\frac{1}{2}\nabla^{2}+V(\textbf{r})+g|\psi(\textbf{r},t)|^{2}-\Omega
L_{z}\right]\psi(\textbf{r},t).
\end{split}
\end{equation}
Here, we allow the possibility to impose a global rotation with frequency $\Omega$
about the $z$-axis, which couples to the $z$-component of the 
angular momentum $L_{z}=-i(x\partial_{y}-y\partial_{x})$. 
The interaction strength $g$ is the two-body interaction in the 2D plane,
which can be varied to a large extent by either tuning the three-dimensional 
scattering length through a magnetic Feshbach resonance~\cite{copyMFR} or 
by changing the $z$-axis confinement through a confinement-induced
resonance.~\cite{copyCR1,copyCR2,copyCR3,copyInt1,copyInt2,copyInt3}

%%%%%%%%%%
\section{Matter-wave interference and Transport Properties}
\label{sec:transport}

%%%%%%%%%%%%%%%%%%%%%%%%%%%%%%%%%%%%%%%%%%%%%%%%%%%%%%%%%%%%%%%%%%%
\begin{figure}[tbp]
\begin{center}
\rotatebox{0}{\resizebox *{8.0cm}{4.0cm} {\includegraphics
{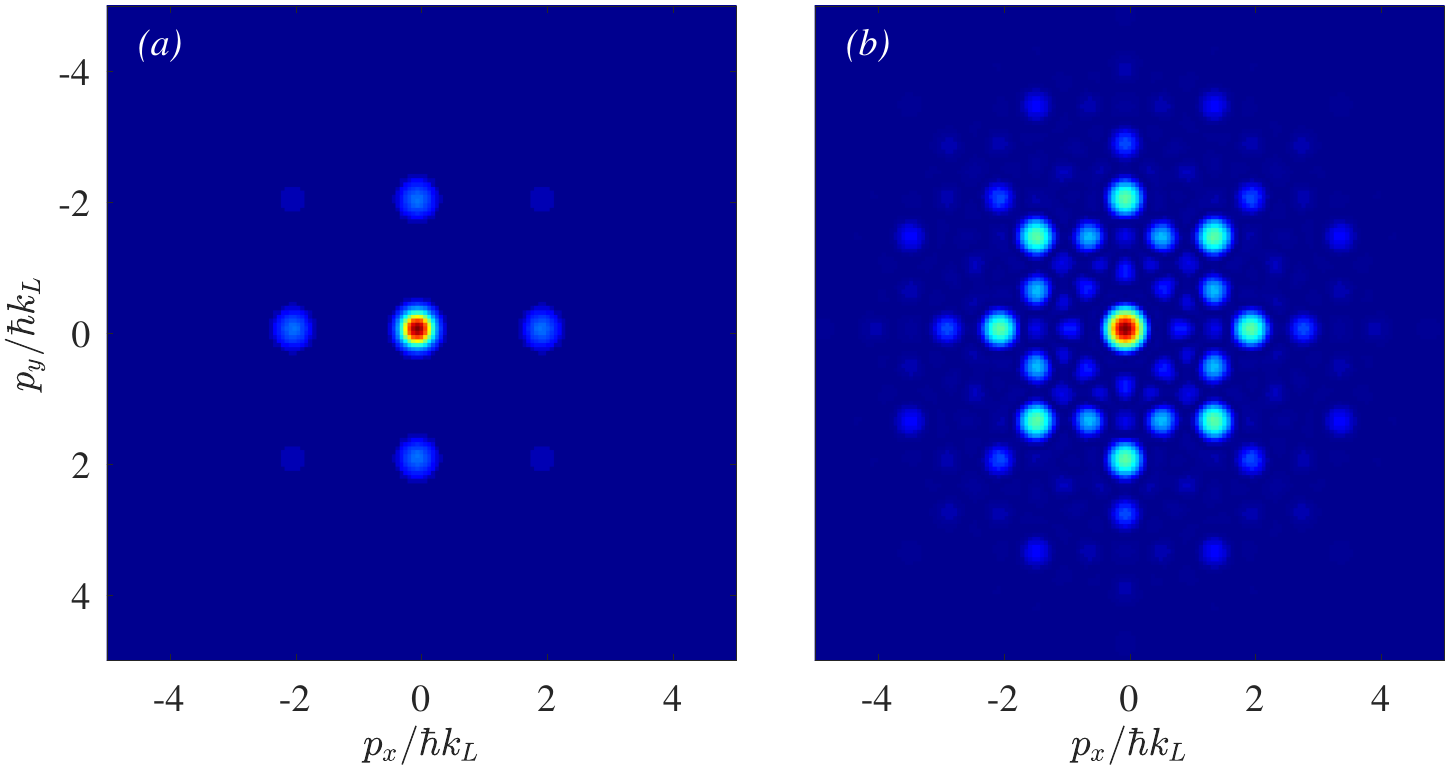}}}
\end{center}
\caption{(Color online) Matter-wave interference patterns of
BECs released from a combined potential of OL and harmonic trap for
(a) a 2D cubic OL with $\varepsilon_{1}=\varepsilon_{3}=1$ and 
$\varepsilon_{2}=\varepsilon_{4}=0$, and (b) a quasi-periodic OL with $\varepsilon_{i}=1$.
Other parameters used in this figure are $V_{0}=-0.2$, $g=10$, $\omega=8\times10^{-2}$, 
and $\Omega=0$.} \label{fig:equilibrium}
\end{figure}
%%%%%%%%%%%%%%%%%%%%%%%%%%%%%%%%%%%%%%%%%%%%%%%%%%%%%%%%%%%%%%%%%%%

In this section, we first analyze the case without global rotation, i.e., $\Omega = 0$.
One typical experiment to reveal the quasi-periodic lattice structure 
is the matter-wave interference pattern in the momentum space extracted from
time-of-flight images. Figure~\ref{fig:equilibrium} depicts the simulated 
results of BECs released from a combined potential of harmonic trap and OL, 
where the initial state is assumed to be the equilibrium wave function $\psi_{0}$ 
obtained for $V_{0}=-0.2$ and $g=10$. Coherence peaks can be clearly observed 
in the time-of-flight images, indicating the indicating the existence of a long-range order.
The positions of peaks correspond to linear combinations of elementary basis
vectors of the reciprocal lattice. For the case of a 2D cubic lattice 
($\varepsilon_{1}=\varepsilon_{3}=1, \varepsilon_{2}=\varepsilon_{4}=0$) 
as shown in Fig.~\ref{fig:equilibrium}(a), the distribution of peaks
displays a clear fourfold symmetry associated with the periodic long-rang order.
For the case of quasicrlystal ($\varepsilon_{i}=1$) as shown Fig.~\ref{fig:equilibrium}(b), 
the distribution of peaks exhibits a clear eightfold rotational symmetry. 
Contrasting to the periodic case, interference peaks are present at momenta 
with smaller magnitude inside the first Brillouin zone, revealing the 
distinctive self-similar fractal structure of quasicrystals.~\cite{copyEight}

%%%%%%%%%%%%%%%%%%%%%%%%%%%%%%%%%%%%%%%%%%%%%%%%%%%%%%%%%%%%%%%%%%%
\begin{figure}[tbp]
\begin{center}
\rotatebox{0}{\resizebox *{8cm}{7cm} {\includegraphics {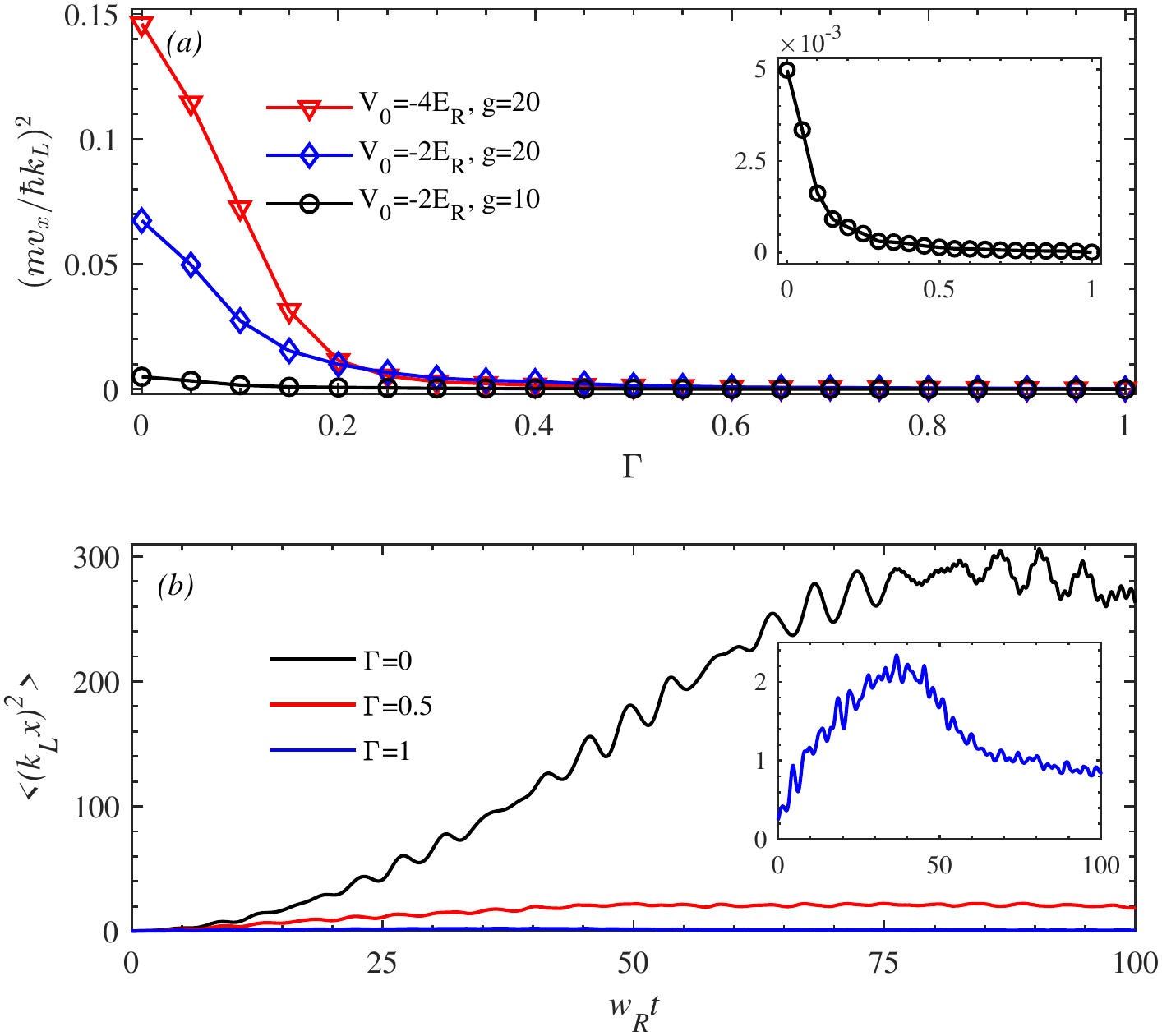}}}
\end{center}
\caption{(Color online) (a) The mean diffusion velocity along the $x$-axis by varying the 
quasi-disorder $\Gamma$ ($\Gamma=\varepsilon_{2}=\varepsilon_{4}$,
$\varepsilon_{1}=\varepsilon_{3}=1$) for different interaction strength $g$ and lattice depth $V_{0}$. 
(b) Time evolution of the mean-square position along the $x$-axis upon released from the 
harmonic trap in periodic ($\Gamma=0$) and quasi-periodic ($\Gamma=0.5, 1$) lattices
with $V_{0}=-2$ and $g=10$. Other parameters used in this figure are 
$\omega = 0.08$ and $\Omega=0$.} \label{fig:dynamic}
\end{figure}
%%%%%%%%%%%%%%%%%%%%%%%%%%%%%%%%%%%%%%%%%%%%%%%%%%%%%%%%%%%%%%%%%%%

Another characteristic feature of quasicrystal is the transport properties subjected
to an external potential gradient. In a periodic lattice, the coherent diffusion of BEC
is frictionless as a result of the off-diagonal long-range order. On the other hand, 
a strong enough disorder would hinder the transport of particles and lead to an 
insulating phase. Thus, one would expect an evolution from diffusive to localized states
by continuously evolving the OL from periodic to quasi-periodic, although a quantitative
description of the process could be very complicated in the presence of 
interaction.~\cite{copyTran1,copyTran2,copyTran3}
To understand the transport properties, we assume the BECs are initially prepared 
in an equilibrium state $\psi_{0}$ in the combined potential of OL and harmonic trap. At time $t=0$,
the harmonic trap is suddenly switched off while the OL is maintained as the background. 
The diffusion of BECs is then determined by numerically solving the 
real-time GP Eq.~(\ref{eqn:3}), and the effect of interaction $g$ and lattice depth $V_{0}$
can be investigated as in Fig.~\ref{fig:dynamic}. Since the quasicrystal structure is 
a superposition of two square lattices tilted by $\pi/2$, the two lattices can be regarded 
as their mutual quasi-disorder. Thus, by varying the intensities $\varepsilon_{2}=\varepsilon_{4}=\Gamma$ 
of one squire lattice while keeping $\varepsilon_{1}=\varepsilon_{3}=1$, we can 
quantitatively analyze the effect of quasi-disorder on coherent diffusion of BECs.

Upon released from the harmonic trap, the BEC tends to expand in space as the 
potential and interaction energies are transformed into kinetic energy. 
At short period of time, the mean-square position of particles along 
the $x$-direction can be approximated by a diffusive behavior $\langle x^2 \rangle \sim v_x^2 t^2$ 
with $v_x$ the diffusion velocity, as shown in Fig.~\ref{fig:dynamic}(b) for typical choices 
of parameters. The expansion gradually ceases and the mean-square position starts
to saturate at long-enough time, since the kinetic energy of particles becomes 
less than the energy difference between adjacent sites of the underlying lattice. 
The expansion of BECs are significantly suppressed by the presence of quasi-disorder, 
while the overall time-dependence remains the same. This behavior is qualitatively 
consistent with the results observed in a fivefold symmetric OL.~\cite{copyQOL2}
Thus, we can extract the diffusion velocity by fitting the mean-square displacement 
at short evolving time, and study the change of $\langle v_{x}\rangle$
versus quasi-disorder $\Gamma$ are shown in Fig.~\ref{fig:dynamic}(a). 
A crossover from ballistic diffusion to spatial localization is clearly identified 
with increasing $\Gamma$ from $0$ to $1$,
during which the lattice configuration evolves from a periodic cubic
lattice to an eightfold symmetric quasi-periodic lattice. Although the diffusion-localization 
crossover is very broad since the system is only mesoscopic with a finite lattice size,
we can still observe from all curves in Fig.~\ref{fig:dynamic}(a) that particles cease
to tunnel when $\Gamma \gtrsim 0.5$.

In the diffusive regime $\Gamma \lesssim 0.5$.
the interaction strength $g$ and overall lattice depth $V_{0}$ can obviously affect the 
diffusion velocity $\langle v_{x}\rangle$. 
When the lattice depth $V_{0}$ of quasi-periodic OL is fixed [see the black-circle and 
the blue-diamond lines in Fig.~\ref{fig:dynamic}(a)], 
a stronger interaction strength induces more potential energy within the system, 
which leads to higher kinetic energy and faster diffusion upon releasing 
from the harmonic trap. On the other hand, a deeper potential well $V_{0}$
with fixed interaction $g$ also enhances diffusion of BECs [see the blue-diamond 
and red-triangle lines in Fig.~\ref{fig:dynamic}(a)]. In fact, since the band width 
becomes narrower with increasing lattice depth, the interaction 
and the harmonic trap become relatively stronger in deeper lattices and both lead to 
faster diffusion.

%%%%%%%%%%%%%%%%%%%%%%%%%%%%%%%%%%%%%%%%%%%%%%%%%%%%%%%%%%%%%%%%%%%
\begin{figure}[tbp]
\begin{center}
\rotatebox{0}{\resizebox *{8.0cm}{5.0cm} {\includegraphics
{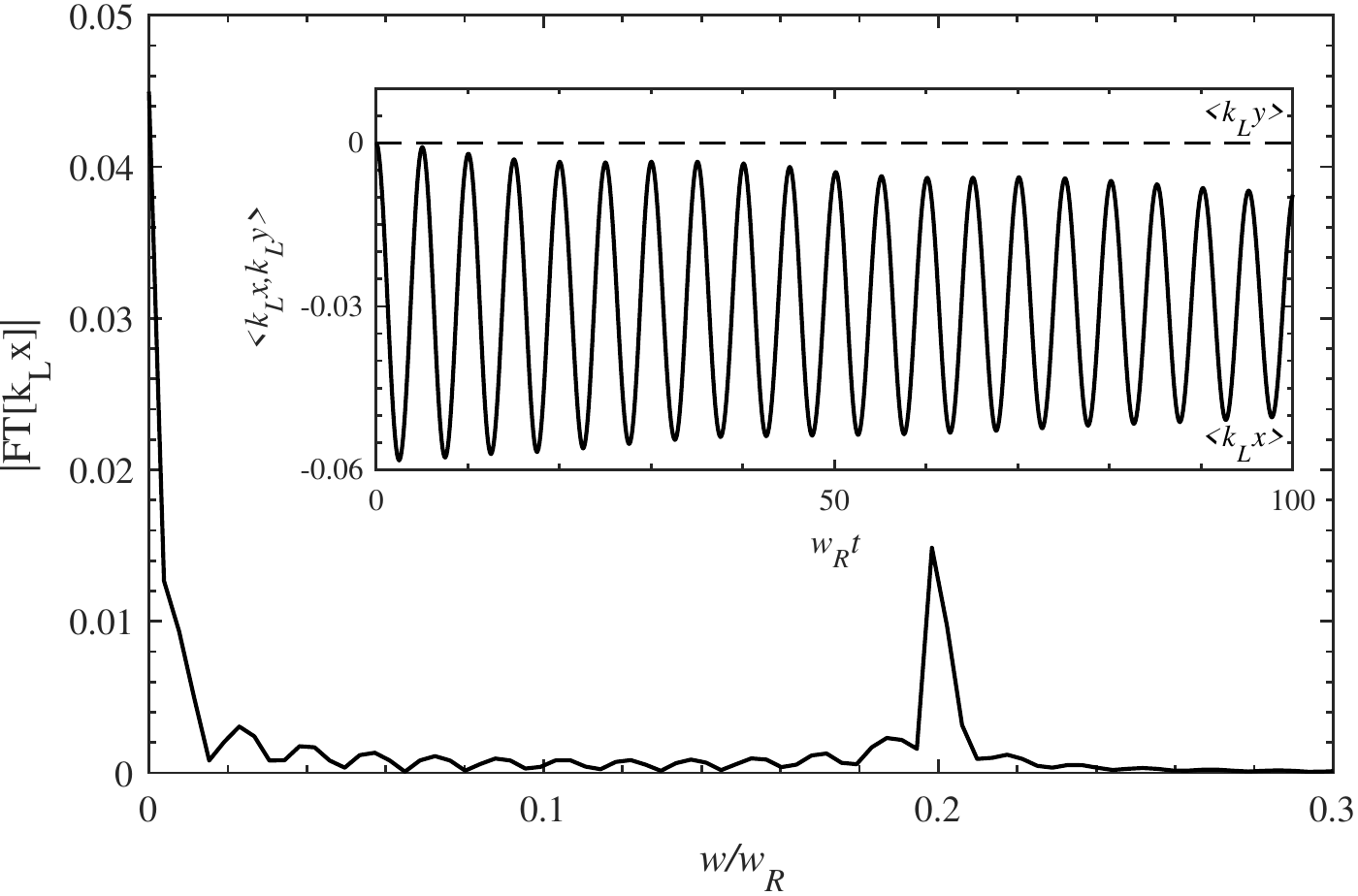}}}
\end{center}
\caption{Fourier transform of the mean position of BECs in the
$x$-direction in the presence of a tilted optical lattice. Inset: Time evolution of the mean 
position of particles. Parameters used in this figure are $V_{0}=-2$,
$V_{\rm tilt}=0.3 k_{L}x$, $g=0$,
$\omega = 0.08$, and $\Omega=0$.}
\label{fig:tilteQOP}
\end{figure}
%%%%%%%%%%%%%%%%%%%%%%%%%%%%%%%%%%%%%%%%%%%%%%%%%%%%%%%%%%%%%%%%%%%

Before concluding this section, we investigate the transport of BECs 
under a static and homogeneous force. 
In periodic OLs, a constant driving force induces oscillatory motion of 
particles known as Bloch oscillation. 
In cold atoms, Bloch oscillation can be observed experimentally by 
applying the constant force via tilt of the optical lattice or 
gravity.~\cite{bendahan-96, wilkinson-96,anderson-98,roati-04}
In the context of quasicrystals, numerical stimulations of 1D tilted 
Fibonacci lattice and 2D tilted Penrose tiling predict a quasi-periodic 
Bloch oscillation.~\cite{copyQOL2,copyFibon} 
Here, we consider a tilting along the $x$-axis of the eightfold symmetric quasi-periodic 
OL with $\Gamma =1$ , and analyze the displacement of particles upon 
time evolution. In order to eliminate the effect of interaction which 
can induce damping and suppression of the Bloch oscillation, 
we assume the BEC is non-interacting with $g = 0$, and the initial state 
is the stationary wave function $\psi_{0}$ for an equilibrium state in 
the combined potential of harmonic trap and OL. At $t=0$, the harmonic 
trap is switched off and the tilting potential is applied, so that the system is 
subjected to a constant force along the $x$-axis. As shown
in the inset of Fig.~\ref{fig:tilteQOP}, particles display quasi-periodic 
oscillation along the $x$-direction, while the motion along the $y$-direction
is completely frozen. The appearance of discrete peaks in the Fourier
transform of mean position $\langle x(t)\rangle$ also verifies an 
ordered structure is present in the quasi-periodic Bloch oscillation,
as depicted in Fig.~\ref{fig:tilteQOP}.

%%%%%%%%%%%%
\section{Vortices and Solitons}
\label{sec:vortices}

%%%%%%%%%%%%%%%%%%%%%%%%%%%%%%%%%%%%%%%%%%%%%%%%%%%%%%%%%%%%%%%%%%%
\begin{figure}[tbp]
\begin{center}
\rotatebox{0}{\resizebox *{8.0cm}{6cm} {\includegraphics {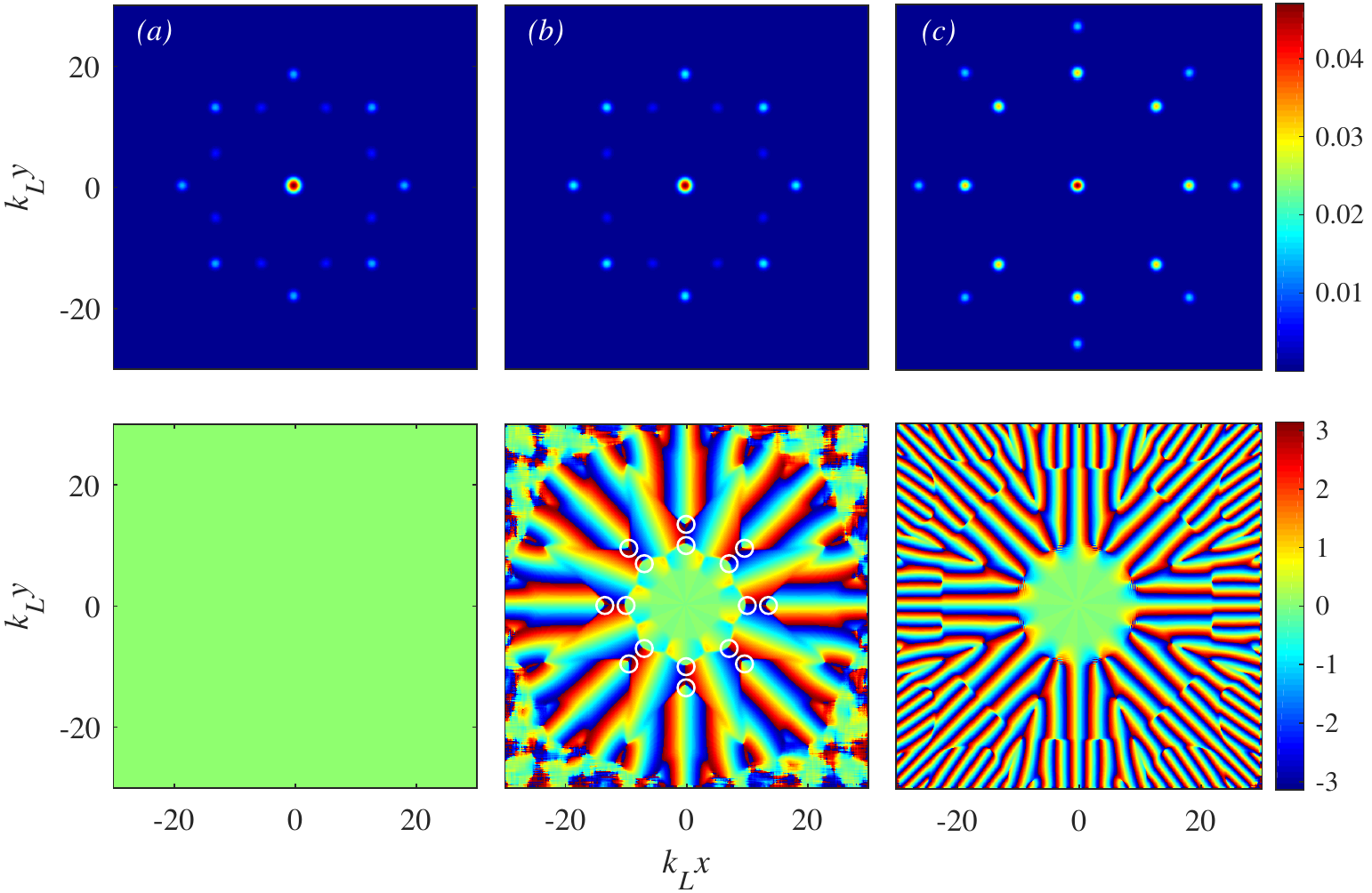}}}
\end{center}
\caption{(Color online) Spatial distributions of particle density $|\psi_{0}|^{2}$ (top row)
and phase $\theta$ (bottom row) of the ground state of a rotating BEC in the combined potential 
of harmonic trap and quasi-periodic OL. The rotating frequency is (a) $\Omega = 0$. (b) $\Omega = 0.05$,
and (c) $\Omega = 0.1$. Other parameters used are $\omega = 0.1$, $V_{0}=-4$, and $g=50$.}
\label{fig:vortice}
\end{figure}
%%%%%%%%%%%%%%%%%%%%%%%%%%%%%%%%%%%%%%%%%%%%%%%%%%%%%%%%%%%%%%%%%%%

Next, we impose an external rotation along the $z$-axis and analyze the steady state 
of BECs in the presence of harmonic trap and quasi-periodic OL.
This configuration attracts great attention as the underlying lattice potential 
can have nontrivial effect on the structure and stability of BEC.
For example, a structural crossover from a triangular to a square lattice 
of vortices with increased potential of square OL has been observed in a 
rotating frame.~\cite{copypin1} It is also predicted that fractional quantum 
Hall effect can present in a dilute gas trapped in a rotating lattice.~\cite{copyHall1} 

%%%%%%%%%%%%%%%%%%%%%%%%%%%%%%%%%%%%%%%%%%%%%%%%%%%%%%%%%%%%%%%%%%%
\begin{figure}[tbp]
\begin{center}
\rotatebox{0}{\resizebox *{8.0cm}{4.0cm} {\includegraphics
{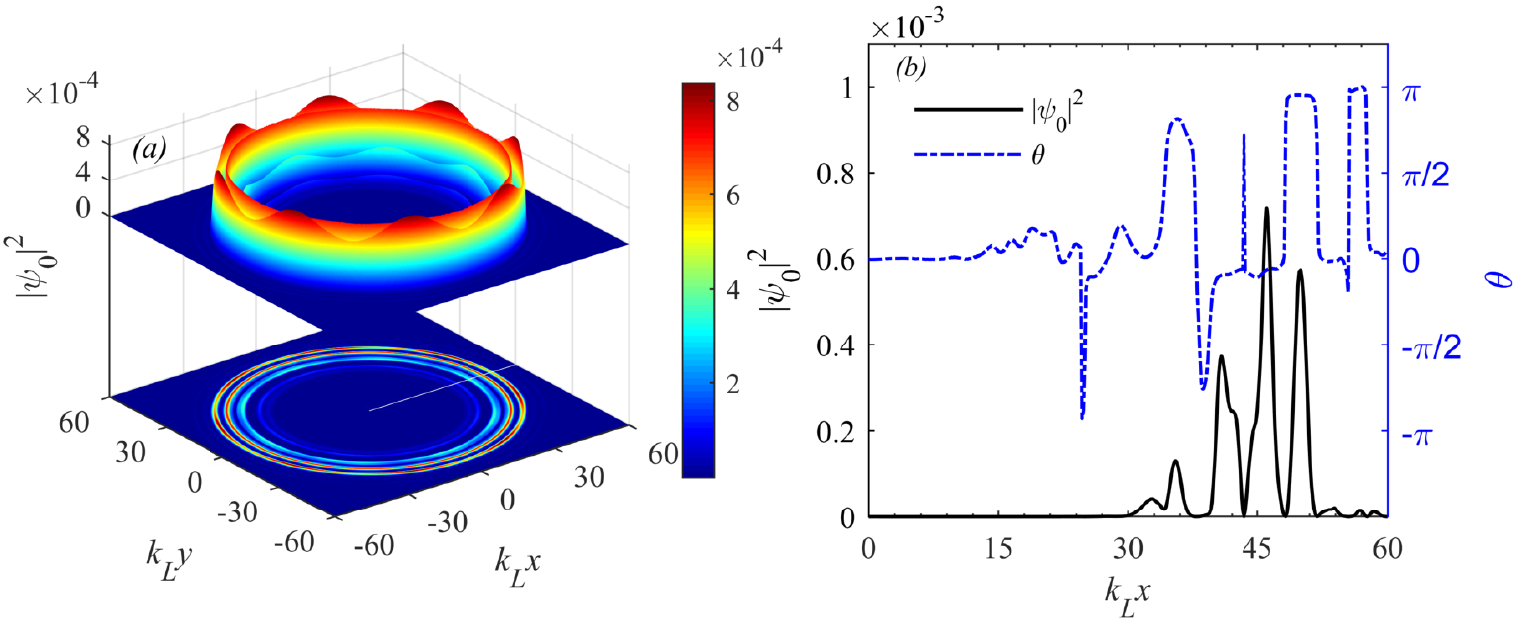}}}
\end{center}
\caption{(Color online) (a) Spatial distribution of particle density $|\psi_{0}|^{2}$ 
in the $x$-$y$ plane. Annular solitons along the radial direction can be clearly 
observed with an approximate axial symmetry. From the surface plot at the top, 
we find that the density distribution acquires a slight axial modulation, reflecting 
the presence of background lattice potential with eightfold rotational symmetry. 
(b) Density and phase profiles along the $x$-direction. Notice that the phase 
(blue dashed-dotted) changes abruptly at the positions of zeros or minima of 
the particle density (black solid). In this plot, the rotating frequency $\Omega= \pi$, 
while other parameters are the same as in Fig.~\ref{fig:vortice}.}
\label{fig:soliton}
\end{figure}
%%%%%%%%%%%%%%%%%%%%%%%%%%%%%%%%%%%%%%%%%%%%%%%%%%%%%%%%%%%%%%%%%%%
Similar method can be implemented to create a rotating quasi-periodic OL to investigate 
the dynamical steady states of BECs therein.
Here we numerically solve for the ground state of BECs in the
combined potential of rotating OL and static harmonic trap,
where the lattice depth $V_{0}$ and the interaction $g$ are fixed. 
When the quasi-periodic OL is stationary with $\Omega=0$ as in
Fig.~\ref{fig:vortice}(a), particles mainly reside at the center of the 
harmonic trap, as well as the lattice sites of lower potential energy.
As a consequence, the spatial distribution of particles preserves the  
eightfold rotational symmetry as the underlying lattice, as one would expect.
The corresponding phase is uniform throughout the entire $x$-$y$ plane
without any singularity. 

By imposing a relatively slow rotation with $\Omega = 0.05\omega_R$ 
and $0.1 \omega_R$ as in Figs.~\ref{fig:vortice}(b) and \ref{fig:vortice}(c), respectively,
the density distribution of particles still preserves the eightfold symmetry with 
dominated probability at the trap center
and lower potential wells. However, the particles are pushed away 
from the center as a result of the centrifugal force. Meanwhile, vortices 
circulating along the same direction of $\Omega$ are generated 
and form an eightfold symmetric lattice structure, as can be identified from 
the phase plots. In the phase plots, the value of phase changes continuously 
from $-\pi$ (blue) to $\pi$ (red). Vortices can be identified at the ends of 
the branch cuts between the phases $-\pi$ and $\pi$, as highlighted by white circles. 

When the external rotation becomes much faster ($\Omega\gtrsim\omega_{R}\simeq10\omega$), 
the vortices decouple from the background lattice in the azimuthal direction, 
and an annular soliton mode appears. As depicted in Fig.~\ref{fig:soliton}(a),
the axial symmetry of soliton is slightly broken owing to the eightfold symmetry of 
the underlying quasi-periodic OL. 
By moving along the $x$-direction from the trap center to the edge, 
we observe multiple sharp phase jumps as shown in Fig.~\ref{fig:soliton}(b).
The positions of these jumps coincide well with the zeros or minima of density 
distribution. For intermediate rotation frequency $\omega \lesssim \Omega \lesssim 10\omega$,
we find that the system is dynamically unstable with no steady solution.
Similar instability has been discussed in a simulation of rotating BECs
in regular periodic lattices.~\cite{copyROL1}

%%%%%%%%%%%%%
\section{Conclusion}
\label{sec:con}

We discuss the properties of Bose-Einstein condensates in an eightfold symmetric 
quasicrystal optical lattice by numerically solving the Gross-Pitaevskii equation.
Starting from an equilibrium state within a combined potential of optical lattice
and harmonic trap, we investigate the matter-wave interference pattern, 
coherent diffusion of particles, and Bloch oscillation subjected to a constant force,
focusing on the characteristic features induced by the quasi-periodic lattice potential. 
In particular, we find that the matter-wave interference pattern can reveal the underlying 
eightfold symmetry of the system, a crossover from diffusive to localized behavior 
can be experienced by continuously tuning the lattice configuration from periodic to 
quasi-periodic, and the Bloch oscillation also acquires a quasi-periodic nature. 
In addition, we also study the emergence and structure of vortices and solitons by 
imposing an overall rotation of the system. For slow rotation, vortices can be generated
and form a lattice structure with eightfold symmetry. For fast rotation, however,
annular solitons would emerge along the radial direction with approximate axial symmetry. 
In the intermediate regime, we find that the system is unstable with no 
steady state solution. Our results provide useful information to the understanding of
the crossover regime between ordered and disordered geometries, and can be 
readily implemented in experiments.

\acknowledgments
This work is supported by the National Natural Science Foundation
of China (Grant Nos. 11434011, 11522436, 11774425), 
the National Key R$\&$D Program of China (Grants No. 2018YFA0306501),
the Beijing Natural Science Foundation (Grant No. Z180013), 
and the Research Funds of Renmin University of China 
(Grants No. 16XNLQ03 and No. 18XNLQ15).
%%%%%%%%%%%

\end{document}